\documentclass[%
 epj,
 nopacs,
]{svjour}

\usepackage{latexsym}
\usepackage{graphics}
\usepackage{graphicx}
\usepackage{amsmath}
\usepackage{footnote}
\usepackage{amssymb}
\usepackage{dcolumn}
\usepackage{newtxtext,newtxmath}
\usepackage{bm}
\usepackage{xcolor}
\usepackage{subfigure}
\usepackage{comment}
\usepackage{float}
\usepackage{url}
\usepackage[utf8]{inputenc}
\usepackage[T1]{fontenc}
\usepackage[ngerman,english]{babel}

\begin{document}
\title{The possibility of twin star solutions in a model based on lattice QCD thermodynamics}
\author{
P.~Jakobus\inst{1,2,3}%
\and A.~Motornenko\inst{2,3}%
\and R.O.~Gomes\inst{3}%
\and J.~Steinheimer\inst{3}%
\and H.~Stoecker\inst{2,3,4}
}  
\institute{School of Physics and Astronomy, Monash University Clayton, Australia. \email{pia.jakobus@monash.edu}%
\and Institut f\"ur Theoretische Physik, Goethe Universit\"at, Frankfurt am Main, Germany%
\and Frankfurt Institute for Advanced Studies, Giersch Science Center, Frankfurt am Main, Germany%
\and GSI Helmholtzzentrum f\"ur Schwerionenforschung GmbH, Darmstadt, Germany}

\date{Received: date / Revised version: date}
\abstract{
The properties of compact stars and in particular the existence of twin star solutions are investigated within an effective model that is constrained by lattice QCD thermodynamics. The model is modified at large baryon densities to incorporate a large variety of scenarios of first order phase transitions to a phase of deconfined quarks. This is achieved by matching two different variants of the bag model equation of state, in order to estimate the role of the Bag model parameters on the appearance of a second family of neutron stars.
The produced sequences of neutron stars are compared with modern constrains on stellar masses, radii, and tidal deformability from astrophysical observations and gravitational wave analyses. It is found that those scenarios in our analysis, in which a third family of stars appeared due to the deconfinement transition, are disfavored from astrophysical constraints.
}

\maketitle
\section{Introduction}
It is presumed that neutron stars (NS) can contain deconfined quark matter due to the high densities achieved in their interiors and might therefore play a decisive role {along with} {both} low {and} high energy nuclear physics in the exploration of the strong interaction and the Quantum Chromodynamics (QCD) phase diagram. Next to particle accelerators~\cite{Bzdak:2019pkr}, neutron stars open an alternative window into the structure of the densest matter in our universe~\cite{Baym:2017whm,Most:2018eaw}.\\
The extremes of QCD matter manifest in similar form in both the stellar phenomena of merging neutron stars and in the laboratory through relativistic heavy ion collisions~\cite{Bauswein:2013yna,Hanauske:2017oxo,Hanauske:2019qgs,Adamczewski-Musch:2019byl}, implying that similar densities and temperatures are excited in rather different physical systems.
One particular possible feature of the QCD phase diagram is of great interest: the possible existence of a first order phase transition from a hadronic phase to a system of deconfined quarks. Such a phase transition can be discovered astrophysically in the properties of compact stars and their corresponding mass-radius relations {from the Tolman–Oppenheimer–Volkoff equation (TOV)~\cite{Tolman:1939jz,Oppenheimer:1939ne}, tidal deformability~\cite{Hinderer:2009ca}, as well as dynamical observables from the binary mergers of neutron stars as in the GW170817 event~\cite{TheLIGOScientific:2017qsa,Most:2018eaw}. 
New methods, such as machine learning~\cite{Fujimoto:2019hxv,Ferreira:2019bny} and Bayesian analysis \cite{Steiner:2017vmg,Raithel:2019uzi,Traversi:2020aaa} are being developed in order to directly extract the equation of state (EoS) from available data.}\\
In context of the QCD phase diagram, astrophysical searches for twin stars or a third family of compact stars in a disconnected stable branch in the mass-radius relation, are of particular interest. Both these cases imply the existence of an isolated branch of stable compact star configurations in the mass radius diagram. The different radii are assumed to be a result of distinct particle compositions in stellar interiors. If twin stars are formed, this would give a direct hint towards a sharp phase transition in QCD matter, e.g. a phase with hadronic degrees of freedom and a second one composed of deconfined quarks. The most prominent scenario is a realization of three families of compact stars: white dwarfs, neutrons stars and their stable twins~\cite{Glendenning:1998ag,Schertler:2000xq,Dexheimer:2014pea,Alford:2015gna,Zacchi:2016tjw,Drago:2017bnf,Alvarez-Castillo:2018pve,Gomes:2018bpw}.
Recently the concept of a ``delayed phase transition`` has been proposed, in which a metastable hypermassive star, developed some time after the
merger event, exhibits a quark core. Two distinct post-merger gravitational-wave frequencies, before and after the phase transition could as well be a promising signature for the existence of quark matter~\cite{Weih:2019xvw,Hanauske:2020tbp}.\\
The EoS for compact stars is often calculated on a basis of nuclear interaction models~\cite{Serot:1984ey,Glendenning:1991ic}, including additional hyperonic degrees of freedom~\cite{Dexheimer:2008ax,Weissenborn:2011ut,Gomes:2014aka,Chatterjee:2015pua,Tolos:2017lgv} and models based on quark interactions~\cite{Nambu:1961fr,Chodos:1974je,Lattimer:2006xb}, to name a few.
In this work we will take a slightly different approach, by employing a quark hybrid EoS in which the model parameters are not only fixed to known nuclear matter properties, but also describe the smooth deconfinement transition and thermodynamics at high temperatures and vanishing densities, obtained from state-of-the-art lattice QCD calculations~\cite{Steinheimer:2011ea,Motornenko:2019arp}. \\
To account for a possibility of twin star solutions we modify the model, which usually only contains a crossover transition to a phase with hadrons and quarks, at high densities by constructing a transition to a deconfined phase of quark matter. This new construction allows to study various possible scenarios involving a first order phase transition to deconfined quarks.
We use the SU(3)-flavor parity-doublet Polyakov-loop quark-hadron mean-field model (CMF) to describe a hadronic and quark system in which interactions are driven by mean field meson exchange and repulsive excluded volume interactions. This model is in agreement with the mass-radius and tidal deformability constraints of astrophysical observations and has been used to investigate the properties of compact stars~\cite{Motornenko:2019arp}. Because the high density region of the EoS cannot be constraint by methods of lattice QCD~\cite{Bazavov:2017dus,Vovchenko:2017gkg,Motornenko:2020vqm}, we will modify the model to investigate various possibilities of 1st order phase transitions at large density. Perturbative QCD calculations in this regime suggest that the pressure of QCD matter is below the Stefan Boltzmann limit of massless non-interacting gas of three quark flavors~\cite{Kurkela:2009gj}.
For the high density regime of the quark phase, we test two different models: The MIT Bag model where we vary the Bag parameter $B$ and the Bag model enhanced by vector meson repulsion (vMIT), in which we study different coupling constants $g_\mathrm{q}^\omega$ between quarks and the $\omega$ vector field. The $g_\mathrm{q}^\omega$ defines the strength of repulsive interactions of quarks by a vector $\omega$ meson exchange. 
Even though such a new EoS with explicit quark vector repulsion may violate lattice QCD constraints for QCD matter at vanishing density~\cite{Steinheimer:2014kka}. Density dependent couplings to a repulsive field may cure this problem at finite density region of the QCD phase diagram which is relevant for astrophysics.
\\
Our current approach differs from previous works as we use a chiral mean field model whose parameters are chosen to describe lattice QCD data at vanishing baryon density and combining it with two different Bag models has not been done before. Our results show no evidence for Twin stars which fulfill the 2 solar mass constraint. This hints that we should generally be aware of confirmation bias and cherry picking when looking for Twin star solutions. A lot of Equations of State and their underlying theories have enough parameters to predict Twin star existence, however not too little attention should be paid as well to those well established models that contradict their existence.\\
This paper is structured as follows:
In Sections~\ref{chapt1} and~\ref{chapt2}, we present the CMF and Bag models used in this framework, as well as the Maxwell construction formalism to implement the transition. We present our results and discuss them in context of feasibility of twin star solutions in Section~\ref{chapt3}. We summarize our results in section~\ref{chapt4}. 
\section{Chiral mean field model}\label{chapt1}
\begin{figure}[t]
		\centering
		\includegraphics[width=\linewidth]{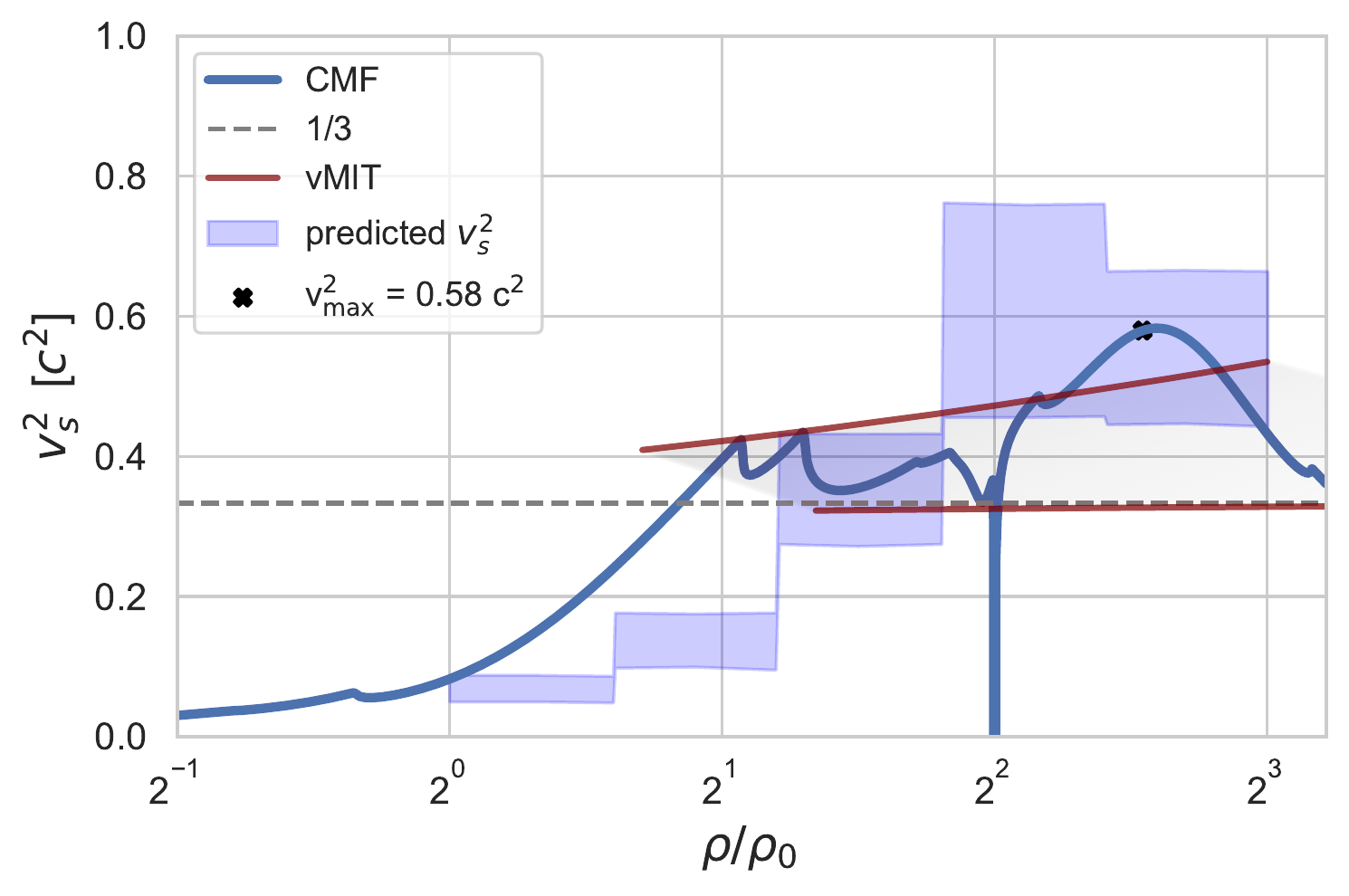}
		\caption{Squared speed of sound  $v_s^2$ for CMF and both Bag models as function of density up to $8\rho_0$ in units of saturation density~$\rho_0$. Shaded area is the estimated $v_s^2$  ($2\sigma$ credible interval) from the deep neural network. Solid blue: CMF model. A first order phase transition at $~4\rho_0$ is reflected in the squared speed of sound by the sudden drop to zero. At this density chiral symmetry is restored. The dashed grey line is the Stefan Boltzmann limit for an ultrarelativistic ideal fluid with squared sound speed $v_\mathrm{s}^2 = 1/3$. The squared speed of sound for the MIT Bag model is $1/3$. The CMF model converges to this value at large densities $\geq 30\rho_B$ (not shown here). The black cross at $\sim 8 \rho/\rho_0$ is the maximum sound speed of the CMF EoS. This regime appears due to the strong repulsion among hadrons. Solid red: Upper and lower limits for vMIT Bag model (using different repulsive couplings).\label{fukus_sos}}
\end{figure}

The Chiral $SU(3)$-flavor parity-doublet Polyakov-loop quark-hadron mean-field model, CMF, describes matter composed of hadrons and quarks. It incorporates several concepts of QCD phenomenology, meson exchange interactions in the baryon octet~\cite{Papazoglou:1998vr}, excluded volume repulsive interactions amongst all hadrons~\cite{Rischke:1991ke,Steinheimer:2010ib}, parity doubling\footnote{Notice that an earlier version of the CMF model does not include the chiral partners of the baryons and it contains a $\Phi$ term in the effective mass of the fermions, which leads to a phase transition due to the deconfinement~\cite{Dexheimer:2009hi}} amongst baryons~\cite{Detar:1988kn} and quarks within a Polyakov loop extended Nambu Jona-Lasinio model~\cite{Fukushima:2003fw}.
The parity doubling assumes that the mass splitting of the baryon masses and their parity partners is generated by scalar mesonic fields, formulated within a mean field approach. As the energy density, (and therefore the scalar density), increases, the mass gap between baryons and their parity partners decreases until degeneracy between the states occurs.
The CMF model includes the full PDG list of hadrons~\cite{Tanabashi:2018oca} which are attributed excluded volume parameters to mimic hadron finite size and their repulsive interactions.
The coupling constants of the hadronic sector are chosen such that properties of nuclear matter are reproduced: ground state density $n_0=0.16\, \mathrm{fm}^{-3}$, binding energy per nucleon is $E_0/B=-15.2$~MeV, asymmetry energy $S_0=31.9$~MeV, and compressibility $K_0=267$~MeV. 
The speed of sound from the CMF model at $T=0$ is shown in Fig.~\ref{fukus_sos} and compared to estimates of a deep neural network analysis, which is based on training data from available mass-radius observations~\cite{Fujimoto:2019hxv}. The CMF model neutron star EoS follows the trend of the $2\sigma$ confidence interval from the neural network at low-moderate densities. However, the uncertainties are still significant and the results of the neural network do not quantitatively constrain the high density regime. 
Note that this model does not naturally exhibit a first order phase transition from nuclear to quark matter, but a 1st order phase transition due to chiral symmetry restoration amongst baryon parity partners~\cite{Motohiro:2015taa}. The abrupt decrease of the speed of sound to the zero value locates the phase transition of the CMF EoS. However this chiral phase transition inhabits a small latent heat, too small to be reflected in the structure of neutron stars~\cite{Motornenko:2019arp} and can be easily hidden in the errorbands of the neutral network analysis. %
According to the results, the CMF model predicts hybrid neutron stars with masses up to~$2\,\mathrm{M}_\odot$. Such stars have a total quark mass fraction up to 30\%. There is no sharp transition in the CMF EoS from hadrons to quarks, so that no second family of quark stars exists within this framework. More details and discussion on this model can be found in~\cite{Motornenko:2019arp}.\\
The CMF model includes scalar-isovector $\rho$-mesons which control isospin asymmetry and are thus relevant for NS matter, where the amount of neutrons is much larger then the amount of protons~\cite{Papazoglou:1998vr,Dexheimer:2008ax,Steinheimer:2011ea,Dexheimer:2012eu}.
The baryon octet couples to the $\omega$, $\rho$ and (hidden strange) $\phi$ field~\cite{Papazoglou:1998vr}.
The baryon masses are dynamically generated by their couplings to the scalar $\sigma$ and strange $\zeta$ field.
These two fields are order parameters for the chiral transition and directly affect the effective baryon masses, see Eq.~\ref{effMass} and~\cite{Steinheimer:2011ea}.
With increasing baryon density $\rho$, the $\sigma$- field decreases and causes the effective masses of the particles to restore chiral symmetry.
The effective masses read
\begin{equation}\label{effMass}
  m^*_{i\pm} 
    = \sqrt{(g^1_{\sigma_i}\sigma 
      + g^1_{\zeta_i}\zeta)^2 
      + (m_0 + n_\mathrm{s} M_{\mathrm{s}})^2}\pm g^2_{\sigma_i}\sigma\pm g^2_{\zeta_i}\zeta,
\end{equation}
where $m_0$ is an explicit mass term of the baryon octet $m_0=759$~MeV, $n_\mathrm{s}$ is the number of strange quarks in baryons and $m_\mathrm{s} = 130$~MeV is the mass of the strange quark. 
The signs~$\pm$ indicate the parity quantum number of the particle.
Finally, $g_{\sigma i}^1$, $g_{\zeta i}^1$, $g_{\sigma i}^2$ and $g_{\zeta i}^2$ are  coupling constants to scalar $\sigma$ and $\zeta$ fields for $i$-th baryon of the octet.
At high densities, quarks are expected to be dominant so a deconfinement mechanism should be incorporated in the model.
This is done in analogy to the Polyakov-loop-extended Nambu Jona-Lasinio (PNJL) model~\cite{Fukushima:2003fw} which is an effective chiral field model for describing quark matter.
The Polyakov-loop $\Phi$ which effectively represents gluon degrees of freedom is controlled by the temperature dependent potential $U(\Phi)$ which is zero for the case of cold neutron star matter~\cite{Motornenko:2019arp}.
The quark masses $m_i^*$ are dynamically generated and controlled by the $\sigma$- and $\zeta$-field.
The effective masses for up, down and strange quarks read
\begin{align}
  m_\mathrm{q}^* &= -g_{\mathrm{q}\sigma}\sigma + \delta m_\mathrm{q} + m_{0\mathrm{q}}, \label{quarkmasses}\\	
  m_\mathrm{s}^* &= -g_{\mathrm{s}\zeta}\zeta + \delta m_\mathrm{s} + m_{0\mathrm{q}},\\
\end{align}%
The $\sigma$-meson controls masses for up and down quarks and the $\zeta$-meson generates the strange quark mass.
The light $u$ and $d$ quarks have the explicit ground state mass term $\delta m_\mathrm{u} = \delta m_\mathrm{d} = 5$~MeV and the heavier strange quark has a mass $\delta_s = 150$~MeV and $m_{0\mathrm{q}} = 235$~MeV.
An additional mass $m_{0\mathrm{q}}$ is introduced to take into account quarks sizable thermal masses which usually appear in EoS models for the quark gluon plasma~\cite{Gorenstein:1995vm,Peshier:1995ty,Levai:1997yx,Vovchenko:2018eod}, this term also prevents quark appearance in nuclear matter.
An explicit volume term $v_\mathrm{B}=v$ is added to the hadrons to suppress them in the quark phase~\cite{Steinheimer:2011ea}.
Consequently, as soon as quarks contribute to the pressure $P$, they suppress hadrons by lowering their chemical potential. For neutron star matter, leptons are taken into account in order to obey charge neutrality and $\beta$ equilibrium.
\section{The high density transition}\label{chapt2}
In order to allow a possible phase transition to a fully deconfined system of quarks, the CMF model is matched to different realizations of the Bag Model.
In the following, we investigate a transition from the CMF model to two versions of the Bag model: the standard MIT Bag Model for the Quark Gluon plasma, with {massless} and {non interacting} quarks ~\cite{Bogo1968,Chodos:1974je} and the vector MIT Bag model~\cite{Klahn:2015mfa}.
\subsection{Bag Model}
\begin{figure}[h]
	\centering 
		\includegraphics[width=\linewidth]{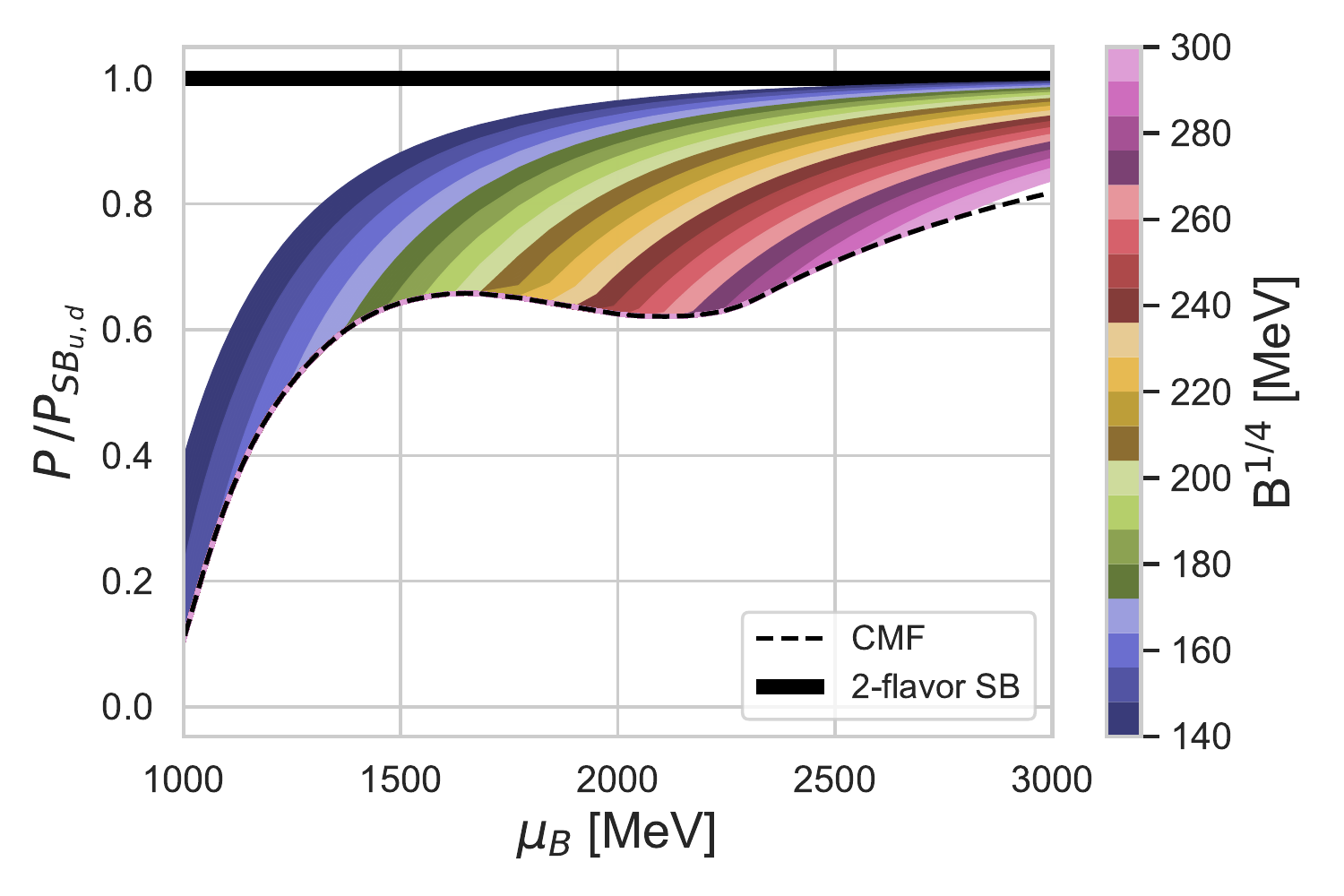}
		\caption{Pressure as function of baryon chemical potential. The colored areas are different Bag model EoS and the black dashed line is the CMF EoS. A higher Bag constant shifts the Bag EoS to the right side and thus leads to a later phase transition.
		The horizontal black line is the Stefan Boltzmann limit.}
		\label{bags}
\end{figure}%
\begin{figure*}[t]
\centering
\includegraphics[width=0.49\textwidth]{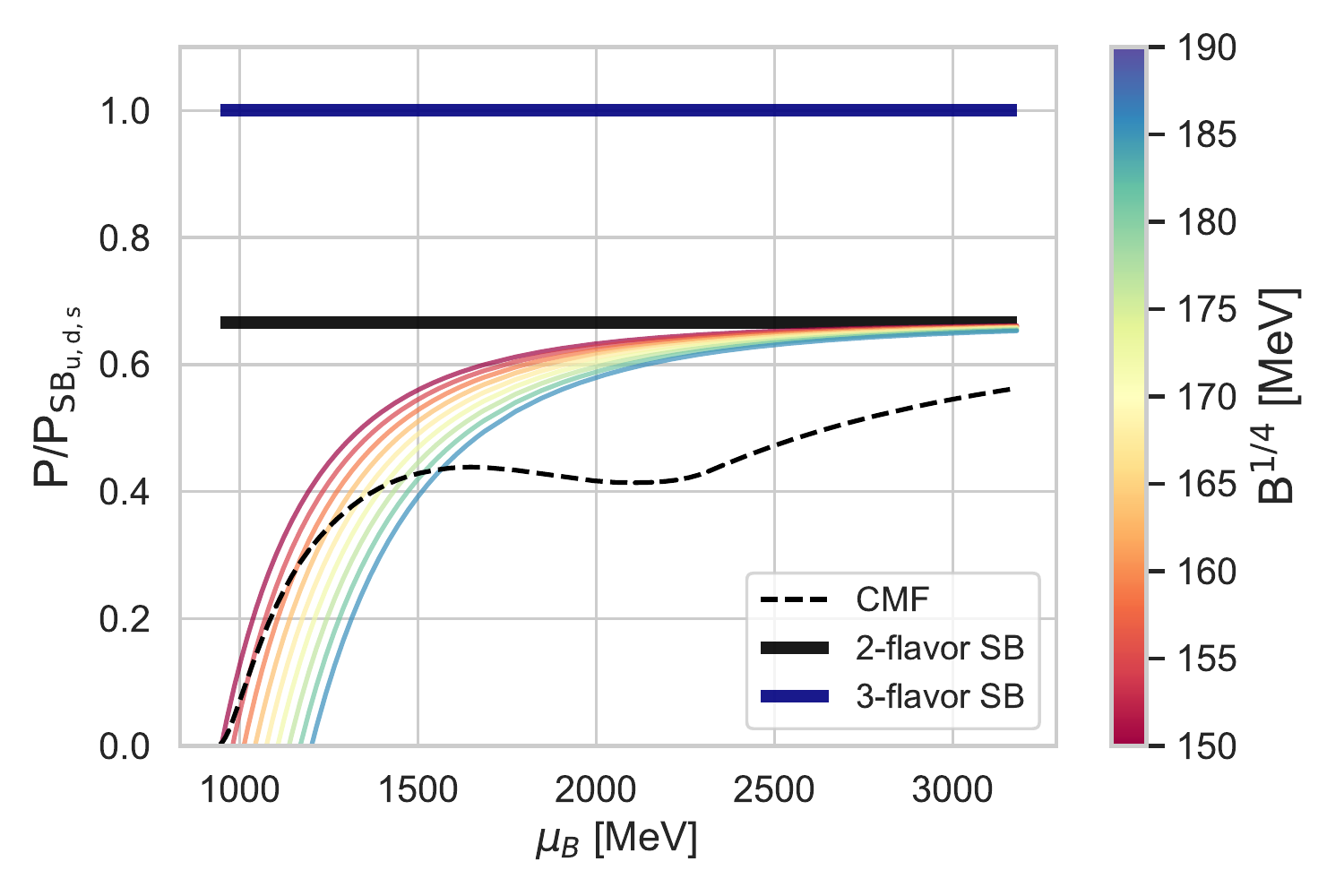}
\includegraphics[width=0.49\textwidth]{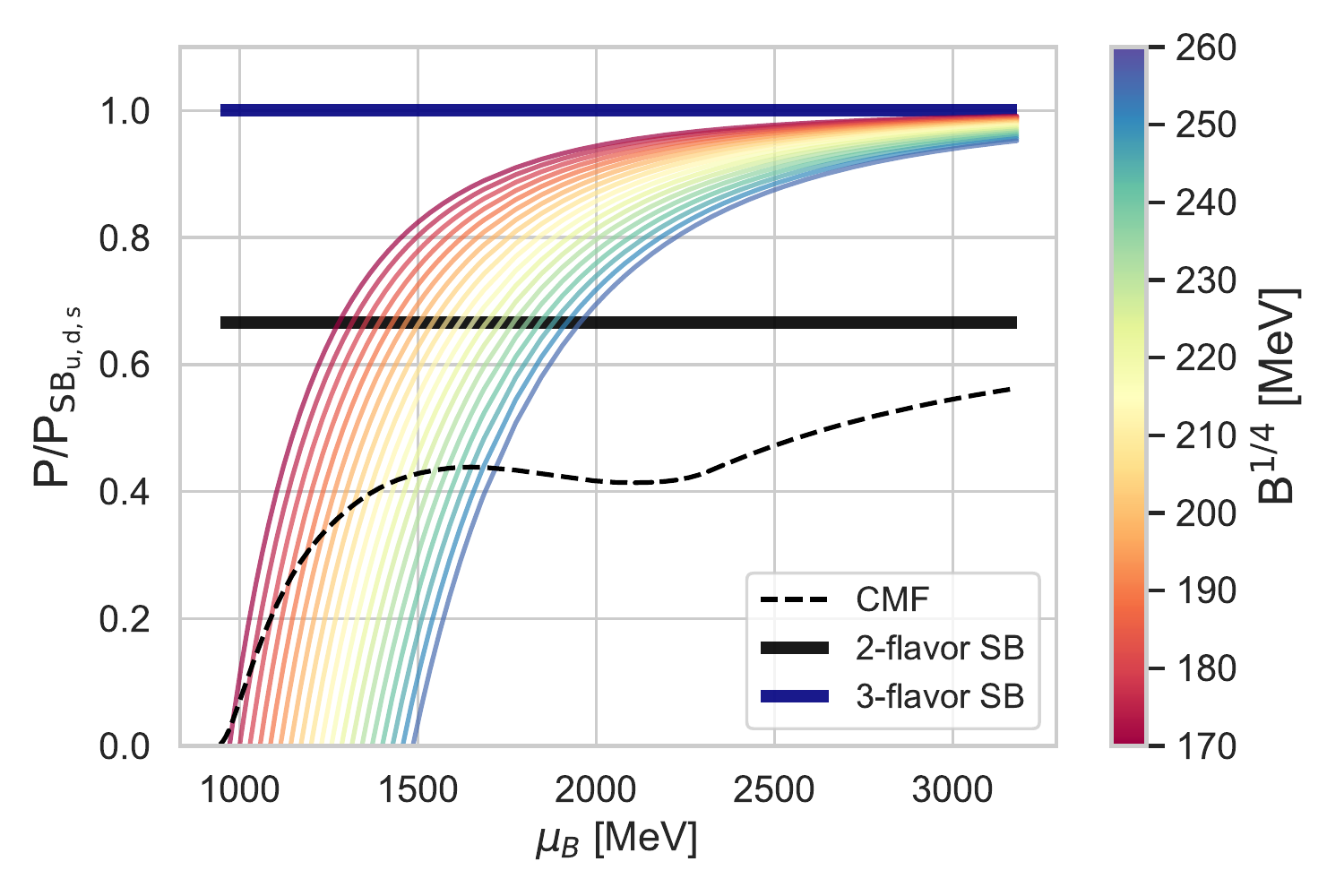}
\caption{Equations of state for the 2-flavor (left) and 3-flavor (right) Bag-CMF model. The dashed black line is the EoS obtained from the CMF model. The colored lines show the MIT Bag EoS with different Bag parameters, see the color code on the right side. The black and blue horizontal lines show the Stefan Boltzmann limit for two and three flavors respectively.}
\label{EoSTwoThree}
\end{figure*}
The first formulation of the Bag model came from Bogoliubov in 1968 who built a theory where three massless quarks inside a spherical volume with radius $R$ are bound in an infinite square well potential~\cite{Bogo1968}\footnote{Unfortunately his paper was only written in french}. Then in 1974, an enhanced Bag model was rediscovered independently~\cite{Chodos:1974je} which then was named the MIT Bag model. In this formalism, quarks are confined in the bag and the surface of the bag has boundary conditions such that the quark current perpendicular to the surface is zero. Vanishing quark current at the boundary of the bag effectively {enforces} quark confinement.
{Several modifications were proposed to the original MIT Bag model, such as the inclusion of medium effects  \cite{Schertler:1996tq}, density  \cite{Bordbar:2012kv} and temperature dependencies \cite{Zhang:2002fs}, and recently it was modified to take into account Hagedorn mass spectra of hadrons~\cite{Vovchenko:2018eod}. Among these modified versions, the ones which include vector interactions among quarks are particularly prominent in the description of massive neutron stars~\cite{Klahn:2015mfa}.}
We relate the baryon and quark chemical potentials by a factor $1/3$, using $\mu_\mathrm{B}=\sum_\mathrm{u,d,s}\mu_{\mathrm{q}_i}$). For such a configuration, the pressure and energy density of the basic Bag Model at zero temperature read:
\begin{align}
  P_\text{QGP} 
    &= \frac{\nu_\mathrm{f}}{4\pi^2} \mu_\mathrm{B}^4 
      - B, \label{finalquant}\\
  \epsilon_\text{QGP} 
    &= \frac{3\cdot\nu_\mathrm{f}}{4\pi^2} \mu_\mathrm{B}^4 
      + B \label{finalquant2}.
\end{align}
For two flavors, the degeneracy factor $\nu_\mathrm{u,d} = 2\times 3\times 2 = 12$ respectively for two spin states, 3 colours and 2 flavors. The three flavor version has a degeneracy factor of $\nu_\mathrm{u,d,s} = 2\times 3\times 3$.
The additional degree of freedom, the strange quark, in the three flavor Bag model may allow (depending on the bag constant) for a bound strange quark matter state~\cite{Farhi:1984qu} where up to $1/3$ of matter is composed of strange quarks which are absent in ordinary matter. Resulting stable exotic nuclear states were conjectured in~\cite{Bodmer:1971we}. The creation of such a type of matter in relativistic nuclear collisions in the laboratory was proposed as signature of quark-gluon plasma formation~\cite{Greiner:1987tg,Greiner:1988pc,Bass:1998vz}, but never confirmed experimentally. 
In the bag model with vector interactions, quarks have non-zero masses and repulsive interaction is taken into account, represented by a coupling $g_V$ of the vector-isoscalar meson $V$ to the quarks. The free leptonic {$e^-$ and $\mu^-$} degrees of freedom are included as well. The modified quark chemical potential of quarks at $T = 0$ reads
\begin{align}\label{potential}
  \mu_\mathrm{q}^* 
    & = \sqrt{k_\mathrm{F,q}^2 + m_\mathrm{q}^2} - g_\mathrm{V} V,
\end{align}
with the Fermi momentum vector $k_\mathrm{F}$ and the bare quark masses $m_\mathrm{q}$. The $\omega$-field suppresses hadronic abundances. The effective chemical potential $\mu_\mathrm{q}^*$ is reduced by the vector interactions.
The phenomenological vMIT Bag model includes chiral symmetry breaking and repulsive vector repulsion~\cite{Klahn:2015mfa} in its Lagrangian.
It has been used in the literature to fulfill two solar mass constraint for neutron star masses from observational astrophysics~\cite{Klahn:2013kga,Klahn:2015mfa,Gomes:2018eiv}.
There are precise Shapiro time delay measurements that observed high NS masses like pulsar PSR J0740+6620 ($2.17^{+0.11}_{-0.10}~\mathrm{M}_{\odot}$)~\cite{Cromartie:2019kug}, PSR J0348+0432 ($2.01\pm 0.04~\mathrm{M}_{\odot}$)~\cite{Antoniadis:2013pzd}, and, recently PSR J0348+0432 ($2.27^{+0.17}_{-0.15}~\mathrm{M}_{\odot}$)~\cite{Linares:2018ppq}.
The vMIT Bag model is an attempt to describe a stiff quark EoS~\footnote{With 'stiff', we refer to the properties of the EoS in their corresponding phases and not to the inevitable 'softening' due to the phase transition.} that supports such high masses. It has an additional term that describes repulsive vector interactions coming from a non-vanishing mean field in the vector meson interaction channel.
However, in the other regime of QCD at high temperatures and vanishing densities, analysis of lattice QCD data disfavors repulsion among quarks~\cite{Steinheimer:2010sp,Steinheimer:2014kka}.
\subsection{Constructing the combined model}
\begin{figure*}[t]
\centering
\includegraphics[width=0.49\textwidth]{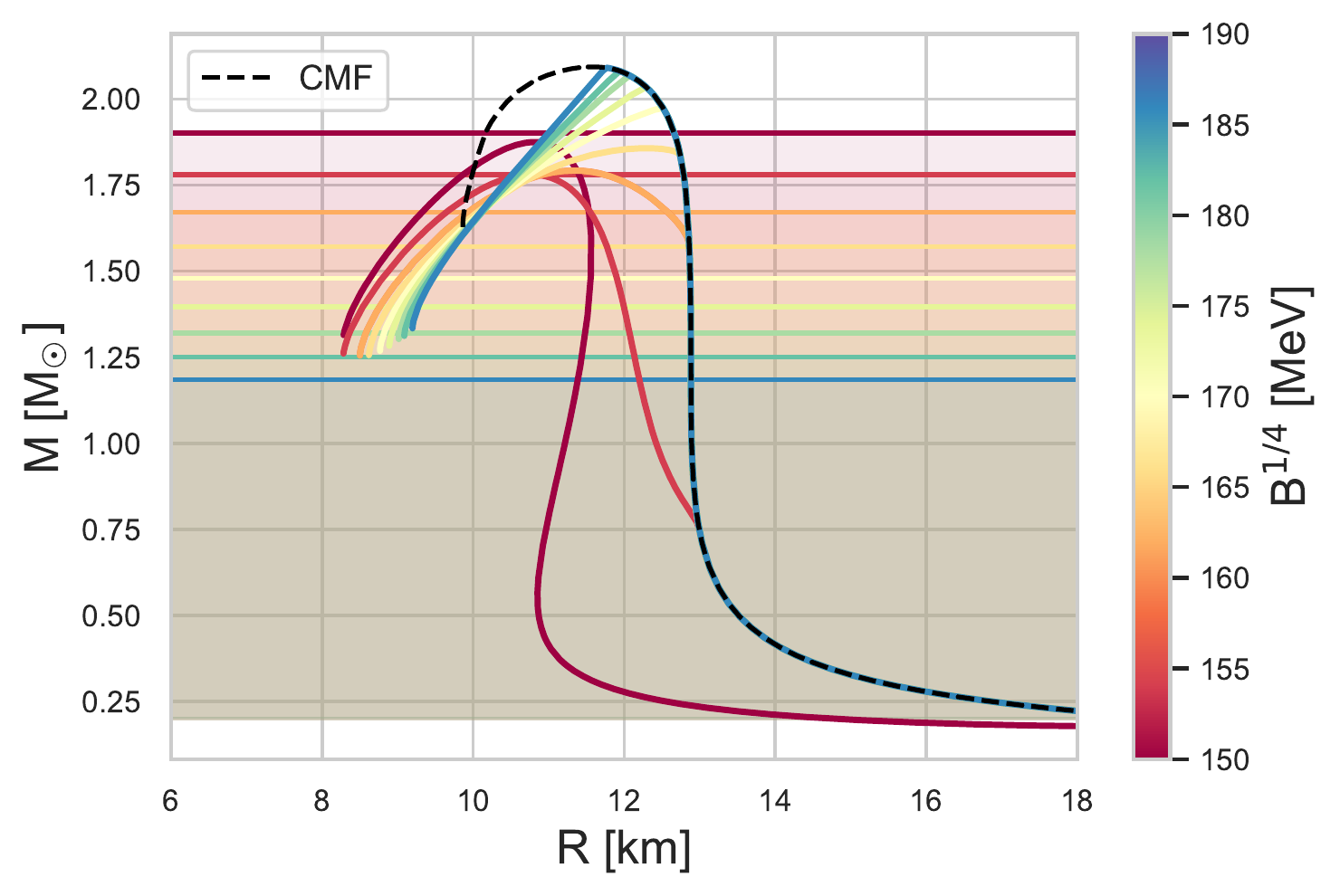}
\includegraphics[width=0.49\textwidth]{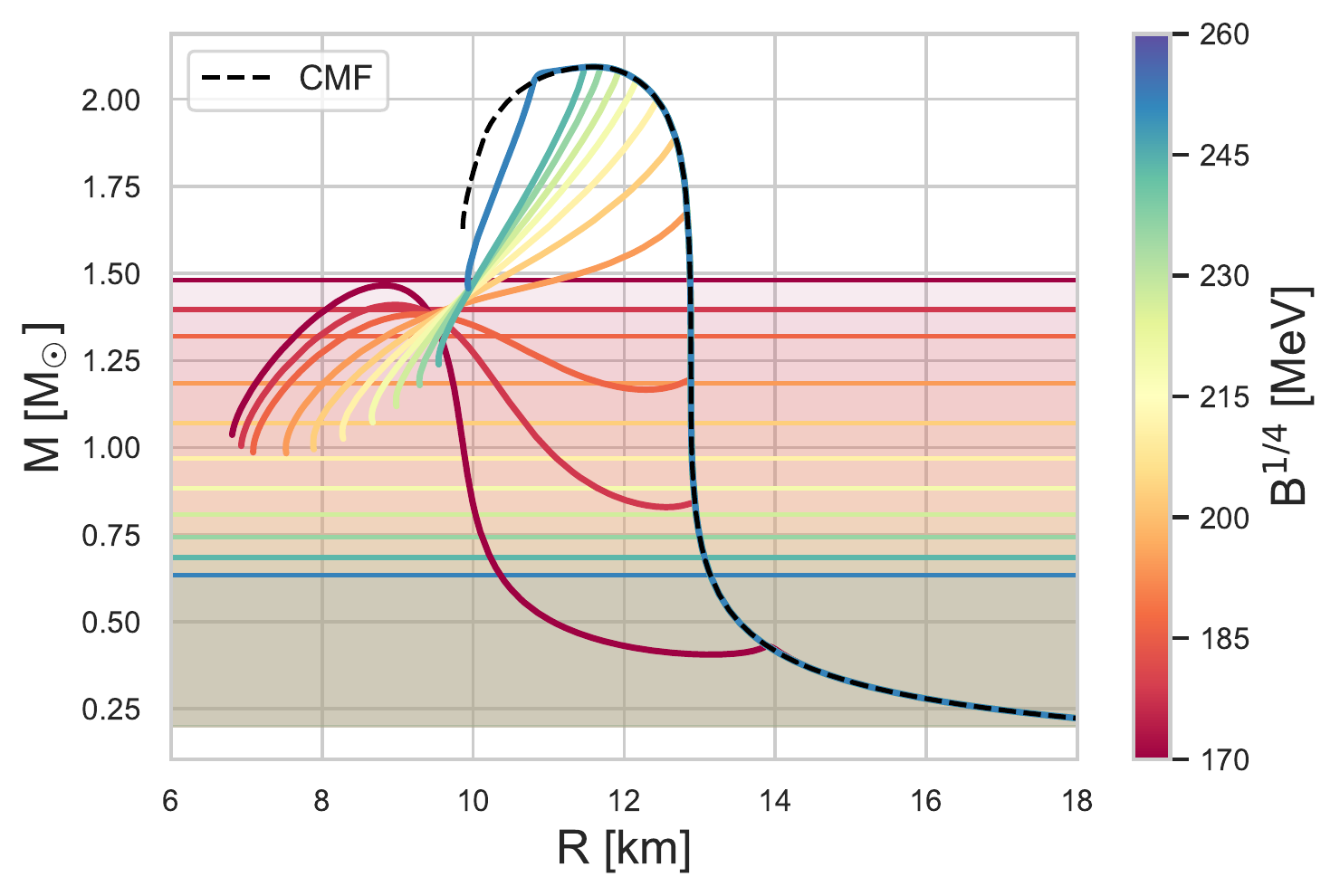}
\caption{
Mass-radius relations for 2-flavor (left) and 3-flavor (right) Bag-CMF, corresponding to the EoSs in Figs.~\ref{EoSTwoThree} (left) and~\ref{EoSTwoThree} (right) respectively. The color indicates values of the Bag constants, as shown in the color code on the right side of the plot. The horizontal colored lines indicate the maximum masses of pure quark stars using the Bag model with respective bag constant as EoS, see Eq.~\ref{bag_mass_limit}, the shaded regions below illustrate the allowed region for pure Bag-matter stars.
The dashed black curve shows the TOV solutions for the CMF model. The combined model ``follows`` this relation up to the point of the Maxwell construction where the branch exits to the left. The combined 2-flavor Bag-CMF model do not produce twin star solutions. For the 3-flavor Bag-CMF model, twin star solutions with masses below 1.5 $\mathrm{M}_\odot$ are predicted.}%
\label{MRTwoThree}
\end{figure*}
We construct a first order phase transition at high baryonic densities to quark matter. The phase transition in our approach is modeled by a Maxwell construction which is well adopted for such a scenario~\cite{Bhattacharyya:2009fg,Montana:2018bkb}\footnote{In the presence of several conserved charges as in the NS matter, e.g. baryonic and electric, non-congruent phase transition occurs~\cite{Greiner:1987tg,Glendenning:1992vb,Hempel:2013tfa,Poberezhnyuk:2018mwt}. In the current version we restrict the construction to the baryonic chemical potential and ignore effects of the electric charge conservation.}. The transition occurs at a point where the two pressures of both EoSs intersect as functions of chemical potential. At the phase coexistence $P_{\rm low}^{\mathrm{cr}} = \sum_i P_{\mathrm{q}_i}^{\mathrm{cr}}$ and $\mu_\mathrm{B}^{\mathrm{cr}} = \sum_i\mu_{\mathrm{q}_i}^{\mathrm{cr}}$, where $P_{\mathrm{q}_i}$ is the pressure contribution of the quarks and $P_{\rm low}$ is the pressure of lower density phase, here -- the CMF EoS.
At the intersection point of both EoSs, baryon number density $n_B$ jumps as well as the energy density $\epsilon$.
{In the following, we refer to the jump in energy density $\Delta\epsilon$ as latent heat, i.g. the discontinuity in energy density at the first order phase transition from the CMF to the Bag model}.\\
Visually, this can be seen in Fig.~\ref{bags}, where the combined EoS corresponds to the maximum at a given baryon chemical potential $\mu_B$ of the dashed black curve (CMF model) and the colored curves (Bag model). The different colors in Fig.~\ref{bags} correspond to different Bag constants in the model, see color code, which we will discuss more in detail in section~\ref{chapt3}. We will vary values of the Bag constant $B$ and additionally, in the case of the vMIT Bag model, the coupling $g_\mathrm{q}^{\omega}$ to study all possible scenarios of the phase transition.
The instability of a star is proportional to the value of $p_{\mathrm{trans}}$ and inverse proportional to the gap $\Delta\epsilon_{\mathrm{trans}}$ in energy density.
Stable twin star branches in our model can only occur if the following condition is fulfilled %
\begin{equation}\label{seidov}
    \frac{\Delta\epsilon_{\mathrm{trans}}}{\epsilon_{\mathrm{trans}}} \geq \frac{1}{2} 
        + \frac{3}{2}\frac{p_{\mathrm{trans}}}{\epsilon_{\mathrm{trans}}}. 
\end{equation}%
This condition is called Seidov limit~\cite{Seidov1971}, it is a generic condition for stellar equilibrium of a star with a phase change. 
It provides a constraint relation between the latent heat $\Delta\epsilon$ and the transition pressure $p_{\mathrm{trans}}$. The constraint is independent of the microscopic model used and the only assumption is, to have a sharp transition (Maxwell construction), which is valid if the surface tension of the interface between the phases is large enough.\\
In case of 3 massless flavors, the pure Bag model itself provides stable quark star solutions with maximal masses that scale as~\cite{Witten:1984rs,Baym:2017whm} %
\begin{equation}\label{bag_mass_limit}
M_{\mathrm{max}}^\mathrm{Bag} \simeq 1.78\left(\frac{155\,\mathrm{MeV}}{B^{1/4}}\right)^2 \mathrm{M}_{\odot}\,.
\end{equation}%
Note, in the present setup there are no Bag matter stars so this relation can not be straightly employed in our calculations since the lower density matter is described with the CMF model.
\section{Results}\label{chapt3}
\subsection{Transition to MIT Bag model}
\begin{figure*}[t]
\centering
\includegraphics[width=0.49\textwidth]{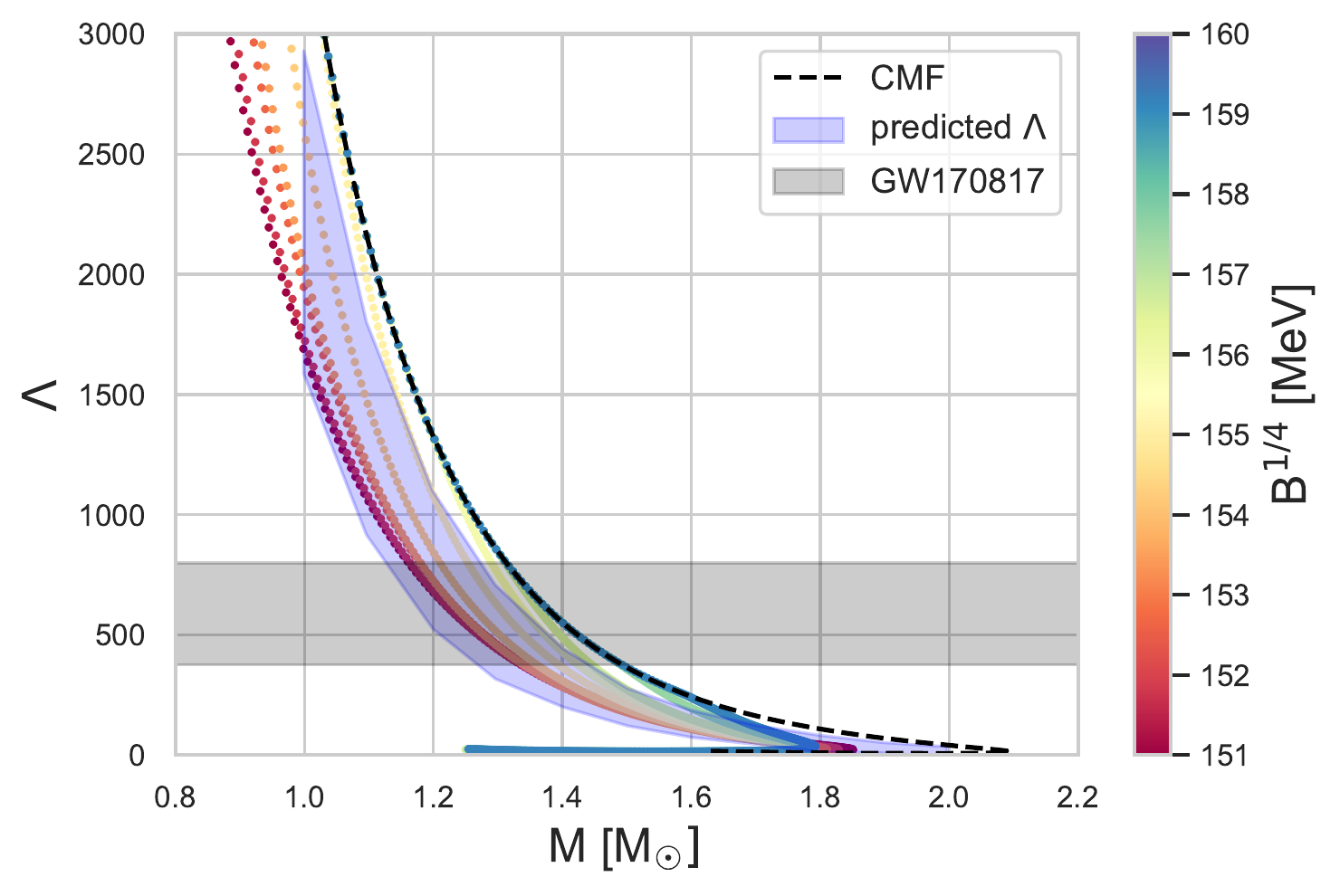}
\includegraphics[width=0.49\textwidth]{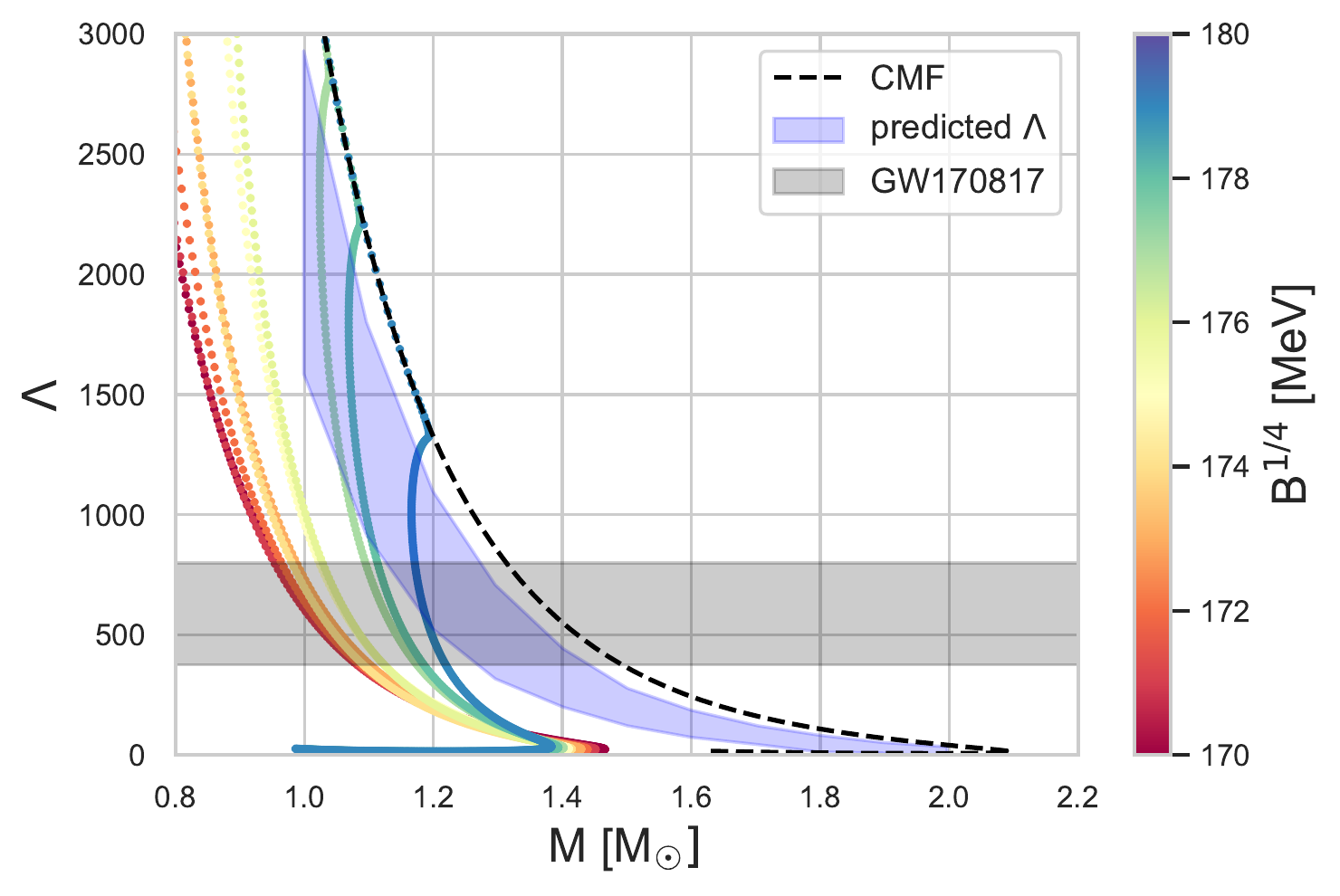}
\caption{Dimensionless tidal deformability parameter $\Lambda$ for the 2-flavor (left) and 3-flavor (right) Bag-CMF model. The blue shaded area is the prediction of a neutral network trained on observed neutron star masses and radii in~\cite{Fujimoto:2019hxv}. The shaded horizontal interval is a $2\sigma$ confidence bound $375 < \Lambda_{1.4\text{M}_\odot} < 800$ from Bayesian analysis based on the merger event GW170817 for a $1.4 \mathrm{M}_\odot$ NS~\cite{TheLIGOScientific:2017qsa}. 
The color indicates the value of the Bag constant, as indicated by the colorbars on the right-hand side. The 3-flavor Bag-CMF model predicts values of $\Lambda$ for stars with masses $\geq 1.2~\mathrm{M}_\odot$ which fall below the constraints on $\Lambda$ from the merger event, as well as below the neural network analysis of NS data~\cite{Fujimoto:2019hxv}.}
\label{deformTwoThree}
\end{figure*}%
In the following we present different mass-radius relations, their corresponding EoS, and the dimensionless tidal deformability $\Lambda$ for our combined CMF-MIT Bag model. Two different scenarios are tested: a combination of the CMF model with a two flavor MIT bag model and the combination of the CMF with a three flavor MIT Bag model, which only differ in the amount of quarks in the Bag model regime. The Bag constant influences the onset of the phase transition and the number of flavors changes the latent heat of the Maxwell construction. Both of these quantities influence possible twin star solutions.\\
Smaller values for the Bag parameter $B^{1/4}$ than 145 MeV are excluded. Otherwise two-flavor quark matter would have a
lower energy than $^{56}$Fe and will form a ground state for ordinary matter different from the one we observe~\cite{Weber:2004kj}.
Starting with the 2-flavor Bag-CMF model model, the pressure in units of $\mu_B^4$ as function of chemical potential $\mu_B$ is presented in Fig.~\ref{EoSTwoThree} (left). 
The combined EoS are constructed such that they, for all $B$, converge to the Stefan Boltzmann limit, see Eq.~\ref{finalquant}. The SB limit for the 3-flavor Bag-CMF model is shifted upwards, see Fig.~\ref{EoSTwoThree} (right). 
The increase of the number of flavors from two to three increases the Stefan Boltzmann limit by 50\%. 
We define the lower limit for the Bag parameter $B$ by requiring that the CMF and Bag models EoS still intersect.
For three flavors, the Bag parameter is shifted upwards to $B^{1/4}\approx 170$~MeV. \\
We observe stable TOV solutions with a significant contribution of Bag matter for Bag constants $B^{1/4} \lesssim 175$~MeV in Fig.~\ref{MRTwoThree} (left). These solutions correspond to the 2-flavor Bag-CMF EoS in Fig.~\ref{EoSTwoThree} (left). %
In Fig.~\ref{MRTwoThree} the horizontal lines indicate the maximum masses which the pure Bag-matter stars can reach, Eq.~\ref{bag_mass_limit}. In the considered combined model, twin star solutions do not appear if the transition occurs above the maximum allowed Bag star mass. 
Stable solutions for the combined model appear for $B^{1/4}\lesssim 160$ MeV, Fig.~\ref{MRTwoThree} (left). 
One can observe that stars above the maximum allowed masses for pure Bag matter stars become consistently unstable near these limiting values, indicated as horizontal lines. But since our stars are composed of not only pure Bag star matter, the point of instability and thus maximum masses deviate slightly.
For the 3-flavor Bag-CMF model in Fig.~\ref{MRTwoThree} (right), we obtain twin star solutions for specific Bag parameters $B^{1/4}\lesssim 200$~MeV with twin star masses $\text{M}\lesssim 1.5~\mathrm{M}_{\odot}$. This contradicts the two solar mass constraint. Higher values of $B$ shift the transition from the CMF to the Bag models to a higher chemical potential and thus higher transition mass in the M-R relation. We can see a correlation regarding the horizontal lines between the maximum allowed pure Bag star masses and a second branch. The dark orange curve for $B^{1/4}\approx 190$ MeV lies, at the onset of transition to the 2nd branch, below the maximum pure Bag star mass of~$\sim 1.3 \text{M}_\odot$ whereas the light orange curve with $B^{1/4}\approx 200$ MeV becomes immediately unstable.
The dimensionless tidal deformability for the two and three flavor model is shown in Figs.~\ref{deformTwoThree}.
The shaded blue area is a result of the neural network analysis of astrophysical observations of neutron stars~\cite{Fujimoto:2019hxv} as in Fig.~\ref{fukus_sos}.%
The tidal deformability for the two flavor Bag-CMF EoS in Fig.~\ref{deformTwoThree} (left) lies within the blue area constraint as well as in the GW170817 merger constraint for $\Lambda$ assuming a mass ratio of 1 with $M_1 = M_2 = 1.4~\mathrm{M}_\odot$~\cite{TheLIGOScientific:2017qsa,Most:2018eaw,Fujimoto:2019hxv}. For three flavors in Fig.~\ref{deformTwoThree} (right) the values for $\Lambda$ lie below both $2\sigma$ confidence intervals, assuming neutron star masses $\geq 1.2 \mathrm{M}_\odot$. NSs with lower tidal deformability are more compact. 
The investigation above demonstrates that, at least for the transition to the CMF-model, the latent heat of the 2-flavor Bag-CMF phase transition is not sufficient to obtain twin star solutions. This is different to previous findings where solutions which allowed for the appearance of a disconnected hybrid branch below the Seidov limit where discussed~\cite{Ranea-Sandoval:2015ldr}.
For the 2-flavor Bag-CMF model, the combination of latent heat $\Delta\epsilon$, transition pressure~$p_\text{trans}$, and transition energy density $\epsilon_\text{trans}$ do not fulfill the condition in Eq.~\ref{seidov}. Thus, either transition pressure/energy density need to be larger/smaller, and/or the latent heat is of considerable extent.\\
Stars with Bag constants $B\geq 180$~MeV become unstable, see Fig.~\ref{MRTwoThree} (left). A reason for that is the upper mass limit for pure Bag model stars, see Eq.~\ref{bag_mass_limit}. %
On the other hand, for a 3-flavor Bag-CMF model twin star solutions occur for lower values of $B$ with masses below $2\mathrm{M}_\odot$. %
A reason for these low twin masses can arise from the nature of the soft MIT Bag EoS with a constant speed of sound $v_s= 1/3$ for any MIT Bag parametrization. 
The first order phase transition has a larger latent heat, compared to our 2-flavor Bag-CMF model. The reason for that is the larger deviation in slopes, that both models each have at the Maxwell construction. This discontinuity leads to a jump in the baryon number density. %
\subsection{vMIT Bag model}
\begin{figure}[h!]
\centering
\includegraphics[width=0.49\textwidth]{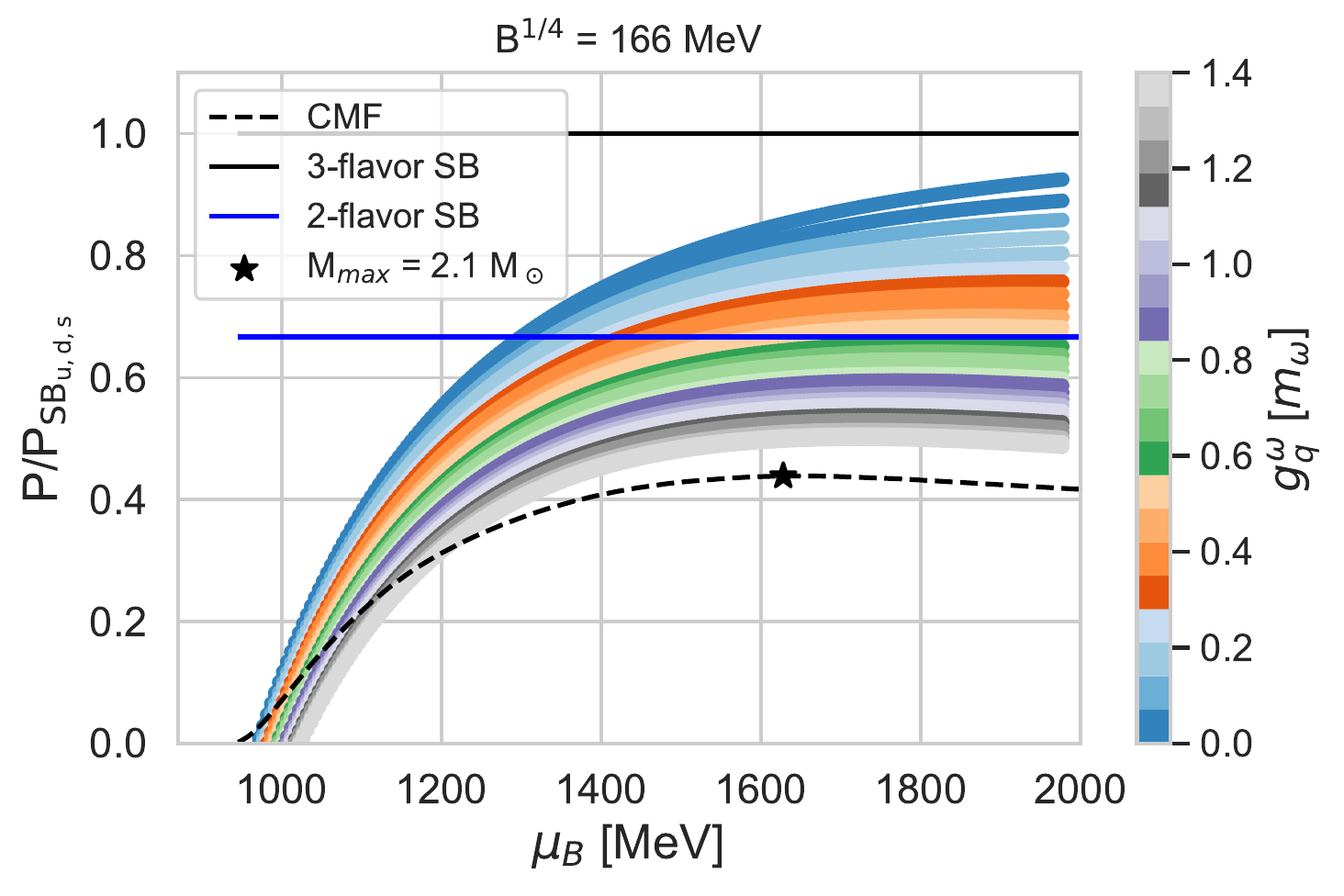}
\includegraphics[width=0.49\textwidth]{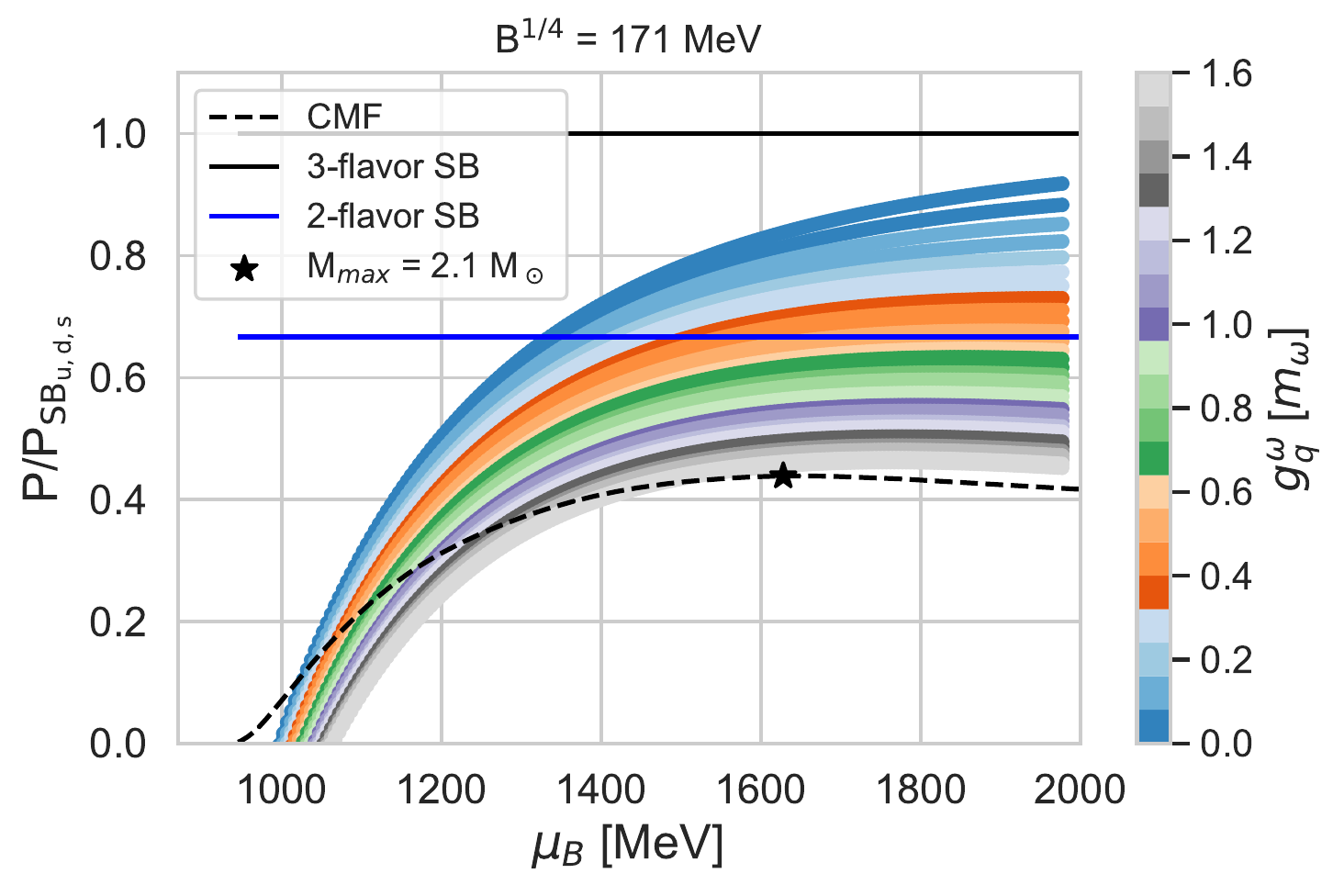}
\includegraphics[width=0.49\textwidth]{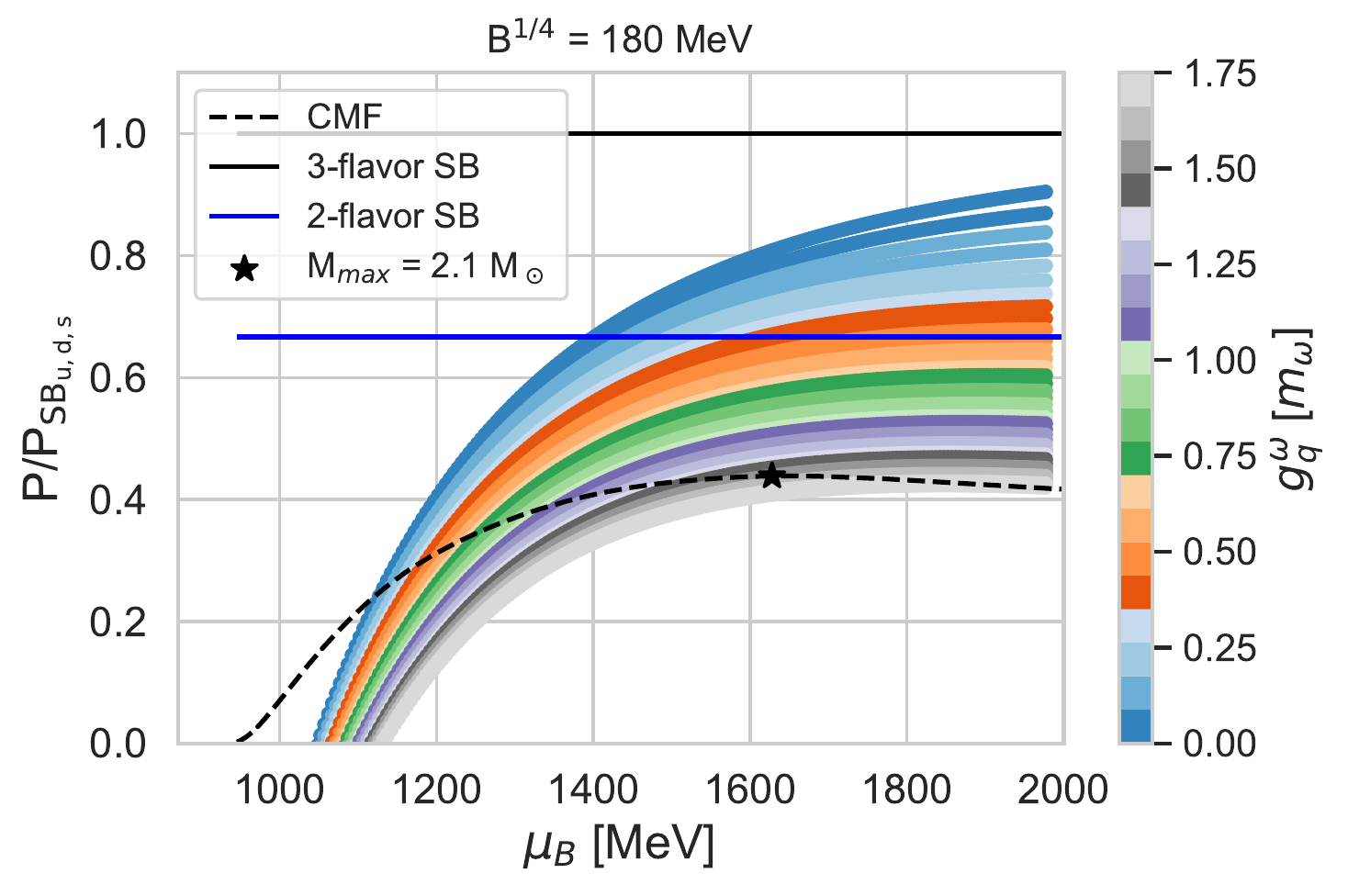}
\caption{3-flavor vMIT-Bag-CMF EoS. The black star $M_{\mathrm{max}}$ is the maximum mass of the pure CMF curve and the blue horizontal line is the two flavor Stefan-Boltzmann limit. Smaller values for $B$ shift the transition to smaller masses. This behaviour was also observed for the pure Bag model. Higher repulsion shifts the transition to a slightly higher chemical potential. It also changes the slope, higher values stiffen the EoS.}
\label{vmit1}
\end{figure}%
\begin{figure}[h!]
	\centering
	\includegraphics[width=\linewidth]{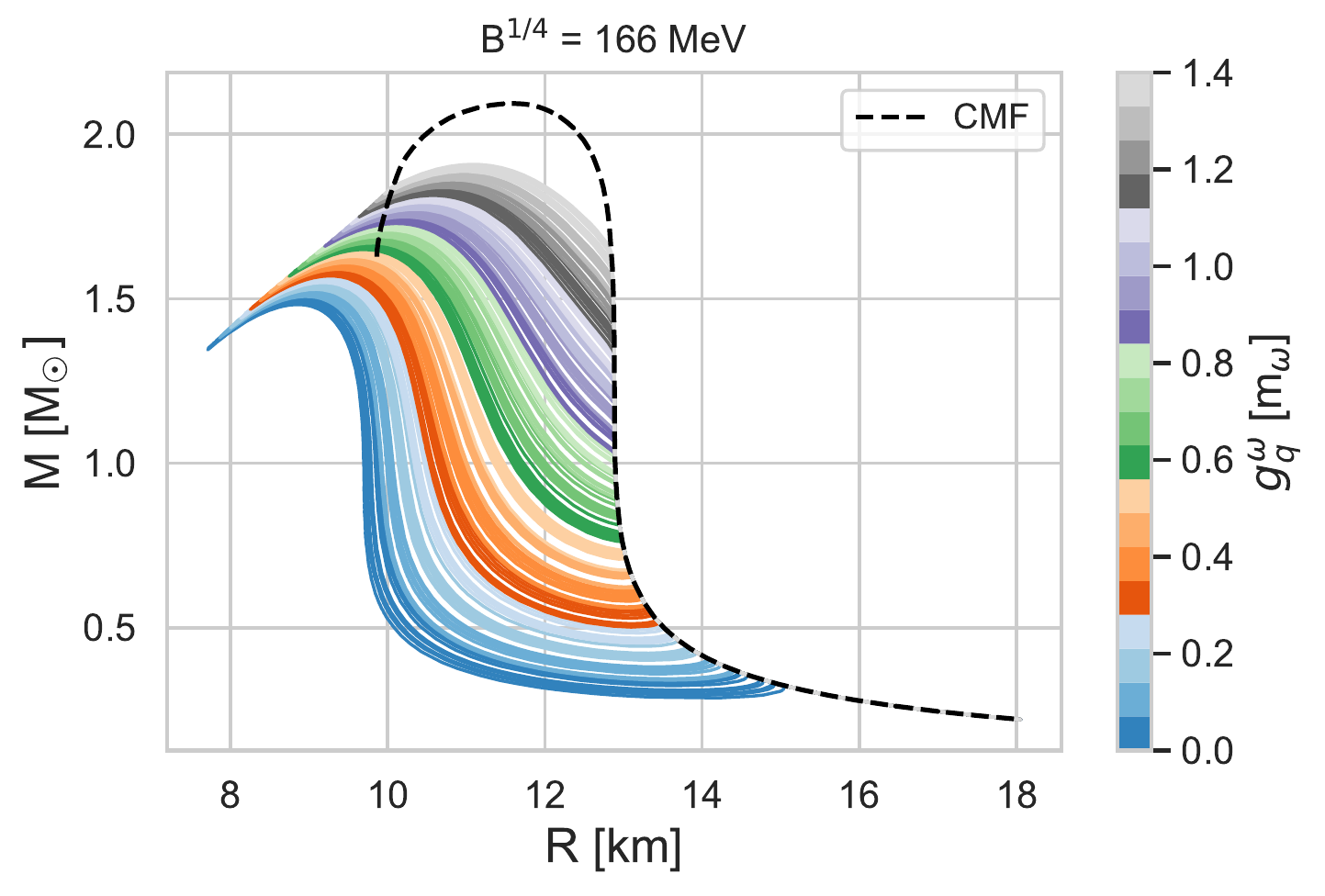}
	\includegraphics[width=\linewidth]{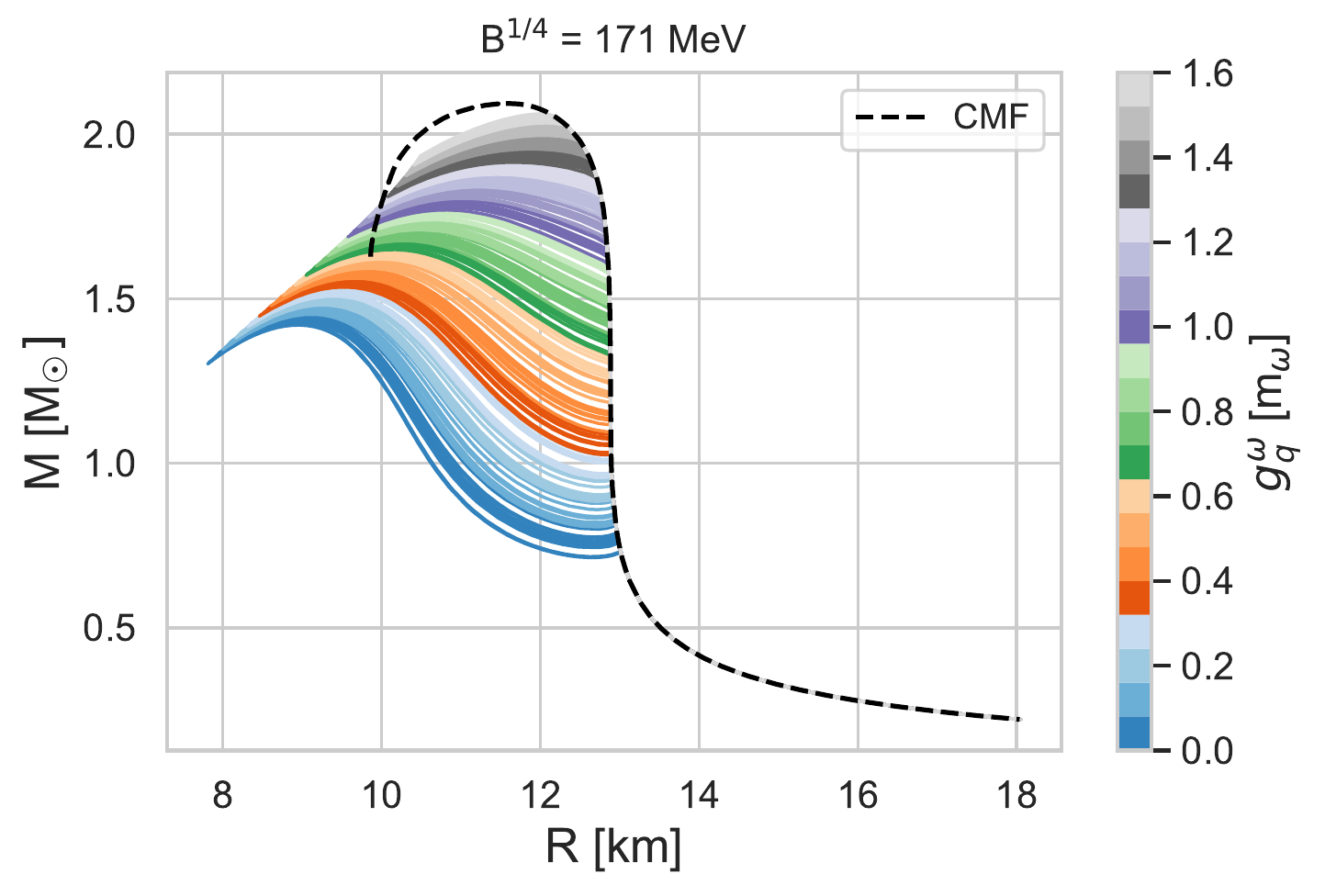}
	\includegraphics[width=\linewidth]{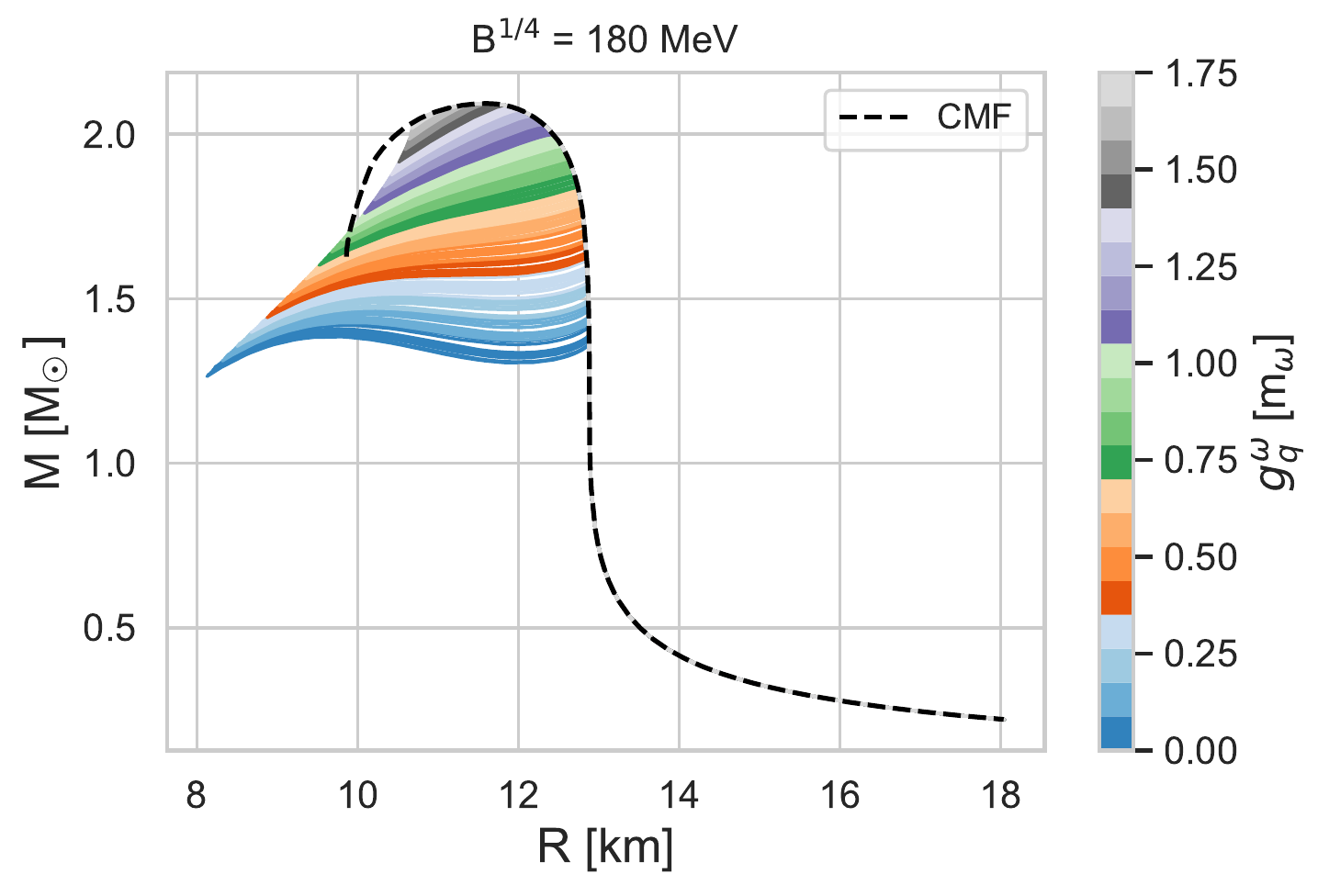}
\caption{M-R relation for three different Bag constants $B^{1/4} \in \{166,171,180\}$~MeV with corresponding EoS in Fig.~\ref{vmit1}. The colorbar in each row shows different quark couplings ranging from small values (blue) to high values (grey). The black star $M_{\mathrm{max}}$ is the maximum mass of the pure CMF curve and the blue horizontal line is the two flavor Stefan-Boltzmann limit. Smaller values for $B$ shift the transition to smaller masses, this behaviour was also observed for both 2- and 3-flavor MIT Bag-CMF models.}
\label{vmitMR}
\end{figure}%
We use a 3-flavor vector enhanced Bag model to construct a stiffer EoS for the quark phase in order to obtain sufficient high masses of $2\mathrm{M}_\odot$ and consider repulsive vector interactions in the quark phase. 
This stiffening could possibly lead to higher mass twin star solutions.
The masses of up, down and strange quarks are $1$~MeV for up and down and $100$~MeV for the strange quark. We vary the coupling constant $g_\mathrm{q}^\omega$ in the range ${g_\mathrm{q}^\omega}/{m_\omega}$=$[0,1.75]$~fm where we use a $\omega$-meson mass $m_\omega = 728$~MeV, see the color code in Fig.~\ref{vmit1} and Fig.~\ref{vmitMR}~\cite{Franzon:2016urz}.\\
Fig.~\ref{vmitMR} and Fig.~\ref{vmit1} show M-R relations and corresponding EoS of 3 different $g_\mathrm{q}^{\omega}$~-~parameterizations where $B^{1/4} \in \{166,171,180\}$ MeV. 
The different color lines with corresponding color code show different quark coupling parameters $g_\mathrm{q}^{\omega}/m_\omega$ in each plot.
$g_\text{q}^\omega$ is the coupling parameter to the $\omega$ field and controls the strength of repulsive force amongst quarks. Increasing the vector repulsion decreases the pressure
at a fixed chemical potential $\mu_B$, hence, an increase of $g_\mathrm{q}^\omega$ will lead to a stiffer EoS~\cite{Baym:2017whm}. 
The black star in each EoS figure is the maximum mass as obtained from the CMF model.
A higher Bag constant leads to a later phase transition, this is the same behaviour as we observed for the MIT Bag model. 
For $B^{1/4} = 180$~MeV we find twins with masses below $1.5~\mathrm{M}_{\odot}$. 
The motivation to choose a Bag model with repulsion was to increase the maximum masses of twins, however it seems that a stiff enough quark EoS alone is not sufficient for stable stars.
In analogy to the reasoning for the flavor variation of the combined CMF MIT Bag model, the quantitative change in slope of CMF and vMIT EoS is a direct measure for the latent heat $\Delta\epsilon$. A higher repulsion eventually leads to a too small jump in the energy density such that the Seidov limit in Eq.~\ref{seidov}, which defines if a star with a specific central pressure can be stable, is not fulfilled and the star is not destabilized by the Maxwell construction.
We summarize the interplay of the coupling strength $a_0$ and the Bag constant $B$ in the framework of the vMIT Bag model as follows: 
\begin{enumerate}
\item The repulsive {coupling $g_\text{q}^\omega/m_\omega$} influences the onset of the transition. A smaller coupling constant leads to a smaller discontinuity in the baryon number density $n_\text{B}$ and thus the latent heat $\Delta\epsilon$. 
\item The {Bag constant} regulates the latent heat. A smaller Bag constant decreases the latent heat and the onset of the phase transition is shifted towards lower chemical potential. Possible twins only occur if the Bag constant is above $\approx 180$~MeV, this softens the {vMIT} {EoS} because it shifts the curve parallel along the x-axis to higher chemical potentials. The star is then immediately unstable after the transition.
\end{enumerate}
Following these points, problems arise when the latent heat $\Delta\epsilon$ is too small. This is the case if the intersection of both {EoS} lie nearly parallel.
A stiff {EoS} in the deconfined phase could help to increase the latent heat.
One way for that to happen could be a larger repulsive coupling constant  $g_\mathrm{q}^\omega$ so that the quark EoS is stiffer than the hadronic EoS.
However, within this framework this is not feasible since $g_\mathrm{q}^\omega$ has an upper limit that arises by requirement of EoS curves to intersect. If the quark EoS is too stiff, then deconfined quark matter always has a higher pressure then nuclear matter, that is not the case in nature since at lower densities hadrons dominate. Having a stiff hadronic {EoS} at the transition, followed by a soft quark {EoS} that stiffens quickly after the transition could possibly lead to twin star solutions. This could be formulated within a density dependent repulsive quark coupling framework, as it has been investigated in~\cite{Soma:2019utv}.
Instead of stiffening the quark {EoS} one could consider to soften the hadronic {EoS} at intermediate densities.
A new analysis of the NICER data gives hint that an extremely soft nuclear EoS and a strong phase transition are mutually exclusive ~\cite{Christian:2019qer}.
A softening of the hadronic phase is possible through the appearance of additional baryonic degrees of freedom. In the CMF model, all hadronic species are included, but at $T=0$ only nucleons and their parity partners appear while other hadronic species are suppressed by their excluded volume-interactions. However, additional softening could result from the appearance of $\Delta$-Baryons or hyperons in the NS EoS due to a decrease of their repulsion or increase of their attractive interactions. The analysis of both lattice QCD data and heavy ion collisions indeed suggest that strange hadrons are subject to smaller EV repulsion due to their smaller size~\cite{Alba:2016hwx,Vovchenko:2017zpj}. We leave the investigation of these systematics for future studies.
\section{Summary}\label{chapt4}
The viability of twin star solutions due to a sharp phase transition to deconfined quark matter was studied in the context of the CMF model. The CMF model is a new type of effective description of QCD thermodynamics which includes the effects of chiral symmetry restoration as well as a coexistence between quarks and hadrons. In addition its parameters where fixed by a matching to lattice QCD thermodynamics. The transition was implemented by a Maxwell construction between the CMF model and different variations of the Bag model.
To investigate different scenarios of the transition, the original MIT Bag model and vector-enhanced Bag model were considered. In the study the parameters of the CMF model remain fixed, but parameters of the Bag model were varied, namely, Bag constant, number of quark flavors and strength of the quark vector repulsion. The variation of the parameters leads the phase transition to occur at different densities and with different latent heat.
In the present framework the mass-radius relations and tidal deformabilities were analysed. 
The observed mass-radius relations suggest that the 2-flavor Bag-CMF model is stiff enough to produce stable configurations with a significant fraction of deconfined quark matter for values of the bag constant $B^{1/4}\leq 175$ MeV. 
However, no twin solutions appear for 2-flavor case in the Bag-CMF construction. For the 3-flavor Bag-CMF model, stable configurations with large quark content appear for bag values $B^{1/4}\leq 175$ MeV, when the latent heat of the phase transition is large enough to destabilize the M-R branch. Howbeit, these solutions only support NS masses up to 1.5$M_\odot$ which is ruled out by astrophysical observations.
The 3-flavor Bag-CMF model was further investigated by an inclusion of the vector repulsion amongst quarks where repulsion coupling was varied as well. The stiffening of the EoS allowed to produce twin star solutions with masses $M\sim 1.3M_\odot$, however higher masses for twin solutions are not supported by this EoS.
Similar to \cite{Ranea-Sandoval:2015ldr} we find that increasing the repulsive vector interaction does not favor a third family branch because both, a higher transition pressure and energy density are more likely to destabilize the quark matter core. Correspondingly, we also observe that if the transition pressure approaches the pressure at which the pure CMF-matter star would become unstable, the hybrid branch tends to be very short, and in our case, immediately unstable (for all our three different Bag model parameterizations) because the denser quark matter core further destabilizes the star. Likewise to \cite{Burgio:2018yix}, we find a large difference in radii for Twin star solutions.  For the 3-flavor Bag model Twin star solutions, the radii lie within $9\, \mathrm{km}\leq R \leq 13\,\mathrm{km}$. The similarities of our findings and mutual constant sound speeds of CSS and MIT Bag approaches propose a comparison of those two models in the future. Contrary to the CSS parametrization with constant sound speed, the vMIT Bag model produces a density dependent sound speed. However despite their deviation in $v_s^2$, our work finds similarities in both approaches. This indicates that a comparison as done in in~\cite{Gomes:2018bpw}, is possible.
The analysis of tidal deformabilities $\Lambda$ and comparison with available constraints also disfavors the suggested scenario of a sharp phase transition to quark matter.
Also the transition to quark matter in NS is constraint by the absence of the vector repulsion among quarks in the low density and high temperature regime of QCD~\cite{Steinheimer:2010sp,Steinheimer:2014kka}.\\
The presented results suggest that stable high mass twin-star solutions may only appear with a density dependent repulsive interaction scheme which incorporates a soft behaviour of quark matter at the density of the Maxwell construction, followed by a stiff quark phase at higher densities. These two characteristics seem necessary in order to obtain a sufficient latent heat at the 1st order phase transition and a second stable branch respectively.
It may be worthwhile to investigate whether such behaviour can be brought in agreement with the measured susceptibilities from lattice QCD simulations which are sensitive probes of density dependent interactions.
    \section*{Acknowledgments}
P. Jakobus and R. O. Gomes would like to thank V. Dexheimer for fruitful discussions and suggestions. The authors thank HIC for FAIR and HGS-HIRe for FAIR. JS thanks the BMBF through the ErUM-Data project for funding and acknowledges the support of the SAMSON AG, WGG-Forderverein, and the C.W. F\"uck-Stiftungs Prize 2018. H. St\"ocker acknowledges the support through the Judah M. Eisenberg Laureatus Chair at Goethe University, and the Walter Greiner Gesellschaft, Frankfurt. Computational resources have been provided by the Center for Scientific Computing (CSC) at the J. W. Goethe-University, Frankfurt.\\
This manuscript is dedicated to the memory of \emph{Prof. Dr. Stefan Schramm}.
\bibliographystyle{spphys.bst}
\bibliography{references.bib}

\end{document}